\documentclass[a4paper,twocolumn,floatfix,preprintnumbers,amsmath,amssymb]{revtex4}

\usepackage{bbm}
\usepackage{amsfonts}
\usepackage{graphicx}
\usepackage{color}
\usepackage{amsmath}
\usepackage{amssymb}
\usepackage{latexsym}
\usepackage{psfrag}

\begin{document}

\title{Piezoresistance in chemically synthesized polypyrrole thin films}
\author{S. Barnoss}
\author{H. Shanak}
\author{T. Heinzel}
\affiliation{Condensed Matter Physics Laboratory, Heinrich-Heine-Universit\"at,
Universit\"atsstr.1, 40225 D\"usseldorf, Germany}
\email{thomas.heinzel@uni-duesseldorf.de}
\author{C. C. Bof Bufon}
\affiliation{Institute for Integrative Nanosciences, IFW Dresden, Helmholtzstr. 20,01069 Dresden, Germany}
\date{\today }

\begin{abstract}
The resistance of chemically synthesized polypyrrole (PPy) thin films is investigated
as a function of the pressure of various gases as well as of the film thickness.
A physical, piezoresistive response is found to coexist with a chemical
response if the gas is chemically active, like, e.g., oxygen. The piezoresistance is studied separately
by exposing the films to the chemically inert gases such as nitrogen and argon.
We observe that the character of the piezoresistive response is a function not only
of the film thickness, but also of the pressure.
Films of a thickness $\lesssim 70\,\mathrm{nm}$ show a decreasing resistance
as pressure is applied,  while for thicker films, the piezoresistance is positive.
Moreover, in some films of thickness $\approx 70\,\mathrm{nm}$, the piezoresistive response
changes from negative to positive as the gas pressure is increased
above $\approx 500\,\mathrm{mbars}$. This behavior is interpreted in terms of  a total piezoresistance which is composed of
a surface and a bulk component, each of which contributes in a characteristic way. These results suggest that in polypyrrole, chemical sensing and piezoresistivity can coexist, which needs to be kept in mind when interpreting resistive responses of such sensors.
\end{abstract}
 \maketitle
\section{\label{sec:1}Introduction}

 One of the most intensely studied semiconductive polymers
 is polypyrrole (PPy) \cite{Chung2005,Vrnata2007,Wang2007,Janata2003,Nagase1993,BofBufon2007,Koezuka1993}.
 This material combines several  advantages. It is stable under ambient conditions when
 highly doped and can be synthesized via both chemical and electrochemical
 polymerization \cite{Kanazawa1980,BofBufon2005}. Moreover, since PPy is piezoresistive and
reacts chemically with various gases,  it has potential applications
in gas and pressure sensors as well as  in actuators.  The detection
of various gases with PPy has been demonstrated, for example $\mathrm{H_2}$ with a resolution limit of $0.06\%$
\cite{AlMashat2008b}, $\mathrm{NH_3}$
\cite{Miasik1986,BinDong2005,Hernandez2007,XingfaMa2008} with a sensitivity of 8 ppm \cite{Carquigny2008},
  $\mathrm{O_2}$ \cite{Kemp1999,BofBufon2006}, $\mathrm{NO_2}$ \cite{Miasik1986,Hanawa1988}with a resolution limit of 40 ppm , $\mathrm{CO_2}$  \cite{Waghuley2008} as well as more complicated molecules like trimethylamine
\cite{XingfaMa2008} or sevofluorane \cite{Wu2007}. Nonmonotonous
evolution to gas exposure as a  function of time has been observed
in several experiments on PPy, see, e.g. Fig. 1 in Ref.
\cite{XingfaMa2008}, Fig. 4 in
Ref. \cite{Miasik1986}, Fig. 3 in Ref. \cite{Hanawa1988}, or Fig. 3 in Ref. \cite{BofBufon2006}. While a
quantitative understanding of the sensing mechanism is still absent,
there are generally accepted suggestions of sensing mechanisms for
some gases. For example, it is assumed that during $\mathrm{NH_3}$ sensing,
an electron is transferred from the ammonia molecule to the polymer,
thereby reducing the hole density and with it the conductivity, as
observed experimentally \cite{Hernandez2007,Miasik1986}. A similar
explanation has been given for $\mathrm{H_2}$ sensing  \cite{AlMashat2008b}, while
during $\mathrm{NO_2}$ or $\mathrm{O_2}$ sensing,  the polymer is oxidized, resulting
in an increased conductivity \cite{Miasik1986}. In some experiments,
however, the response of PPy to the gas was ambiguous. In Ref. \cite{Hanawa1988}, it was observed that for low concentrations of
$\mathrm{NH_3}$, the resistance $R$ increases, while for larger concentrations, a
decrease in $R$ was observed, in contrast to the commonly accepted
sensing mechanism. Furthermore, for some gases, the sensing
mechanism is unclear at present, e.g. for $\mathrm{CO_2}$ \cite{Waghuley2008}
or trimethylamine \cite{XingfaMa2008}.

On the other hand, it has been established that PPy is piezoresistive with a resolution of about $500 \,\mathrm{N/m^2}$ in PPy coated polyurethane foam \cite{Brady2005}. This property has been used in
various demonstrations of PPy based sensors and actuators, e.g. in
wearable sensors \cite{Dunne2005}, as drug delivery valves
\cite{Tsai2005} or as artificial muscles \cite{Fujisue2007}. It is
therefore reasonable to assume that the response of PPy films to gas
exposure may contain both chemical and physical ( i.e., piezoresistive) components.

In the present manuscript, we shed some light on this interrelation by
investigating the response of PPy thin films to exposure of various
gases at different pressures. The studies are carried out with the inert gases nitrogen and argon, as  well as
with the chemically active oxygen. A recently reported technique based on chemical polymerization from the vapor phase \cite{BofBufon2006} is used to prepare comparatively thin PPy films with low roughness. The sensing properties of PPy films prepared by this technique have not been investigated so far. Most importantly, we find that the piezoresistive response changes from negative to positive as the film thickness increases above $\approx 70\,\mathrm{nm}$. Moreover, in some films with thicknesses close to $\approx 70\,\mathrm{nm}$, we observe a change of the polarity of the piezoresistivity (PR) as the pressure increases above $\approx 500\,\mathrm{mbars}$. These findings indicate that the PR in PPy comprises a negative surface component and a positive bulk component. Furthermore, they offer a qualitative explanation for the ambiguous behavior observed in the earlier experiments by Ma et al., \cite{XingfaMa2008},  Miasik et al. \cite{Miasik1986} or Hanawa et al. \cite{Hanawa1988}. The character of the responses furthermore suggest that the piezoresistive surface effect is more suited for sensing applications than the bulk effect.

\section{\label{sec:2}Sample preparation and experimental setup}

\begin{figure}[htb]
\centering
\includegraphics{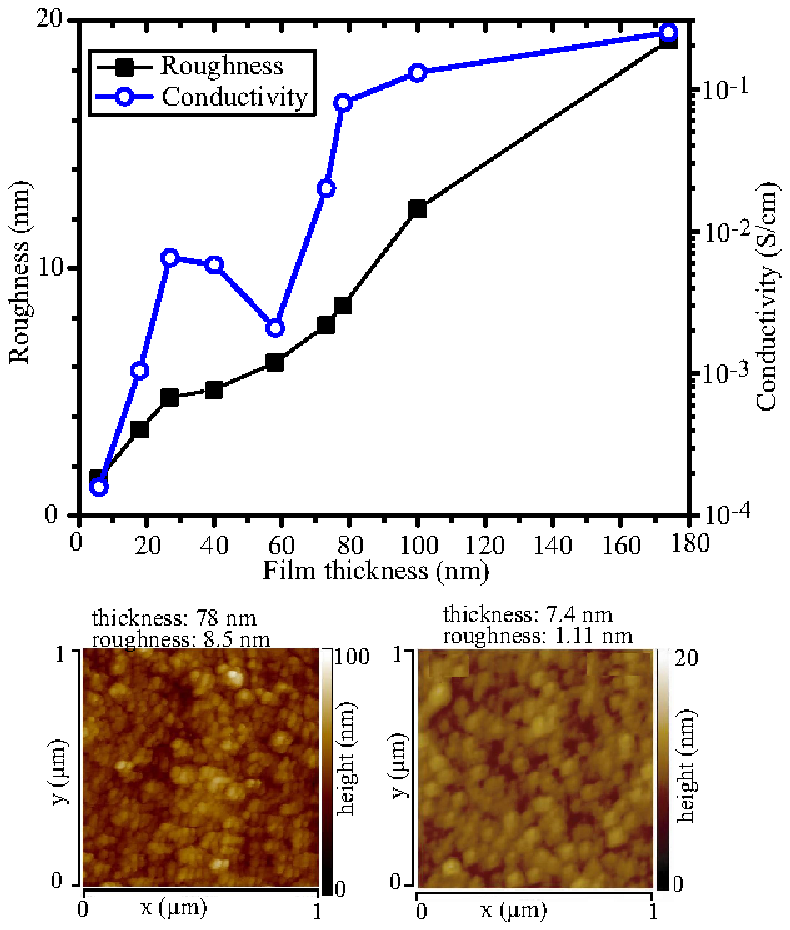}
\caption{ }\label{PPy_Piezoresistivity_Fig1}
\end{figure}

The films are prepared by chemical polymerization of Py monomers from the gas phase.
This technique is described in detail in Ref. \cite{BofBufon2006}. The films are formed
on a glass substrate with Pt electrodes on top which allow 4-terminal measurements. They were patterned by optical lithography on top of the glass substrate and have  a thickness of $100\,\mathrm{nm}$. The separation
between the electrodes is $2\,\mathrm{\mu m}$. The patterned substrate is inserted into
the polymerization chamber filled with Ar gas at pressure of $1\,\mathrm{atm}$.
A droplet of  $\mathrm{H_2O_2:HCl}$ (1000:3 volume fraction) is deposited
on the surface of the patterned substrate. The $\mathrm{H_2O_2}$ serves as
oxidizing agent and the $\mathrm{HCl}$ provides the $\mathrm{Cl^{-}}$ ions which not only are necessary
for charge neutrality, but also act as dopants. The substrate was kept at room temperature
during the polymerization. The pyrrole monomers were evaporated from a boat by
heating it to $100\,\mathrm{^{\circ}C}$. The surface
of typical PPy films are shown in the lower part of
Fig. \ref{PPy_Piezoresistivity_Fig1}, while the roughness and the
conductivity as a function of the film thickness, prepared in one run of sample fabrication, are reproduced in the main figure. A characteristic, egg-like surface morphology is observed, with a roughness of approximately $10\%$ of the film thickness. Such flat films are characteristic for our chemically grown films, \cite{BofBufon2006,BofBufon2007} provided the substrate has a low roughness of the order of $1\,\mathrm{nm}$. The conductivity increases almost three orders of magnitude as the film thickness increases from $6\,\mathrm{nm}$ to $78\,\mathrm{nm}$, and continues to increase weakly as the thickness is further increased. This behavior presumably indicates that highly localized surface or interface effects dominate the transport in the thin films and that bulk conductivity begins to evolve only at thicknesses above $80\,\mathrm{nm}$.

Immediately after growth, the samples are transferred to a vacuum chamber and  vacuum
dried at room temperature for at least one day. The vacuum chamber is equipped with
a hand valve and a pressure sensor (Ionivac Transmitters ITR 90 from Leybold Vakuum) which
allows adjustment and control of  the gas pressure with an accuracy of $\approx 0.5\,\mathrm{mbars}$
for pressures between $50\,\mathrm{mbars}$ and $900\,\mathrm{mbars}$.  Temperature dependent
transport measurements \cite{BofBufon2007} have revealed that the films are p-doped and strongly disordered. The transport is dominated by thermally activated hopping for temperatures above $\approx 30\,\mathrm{K}$, which transforms to Efros-Shklovskii - type hopping at lower temperatures \cite{BofBufon2007}, a typical behavior for strongly disordered semiconductors.

\section{\label{sec:3}Results}

\begin{figure}[h]
\centering
\includegraphics{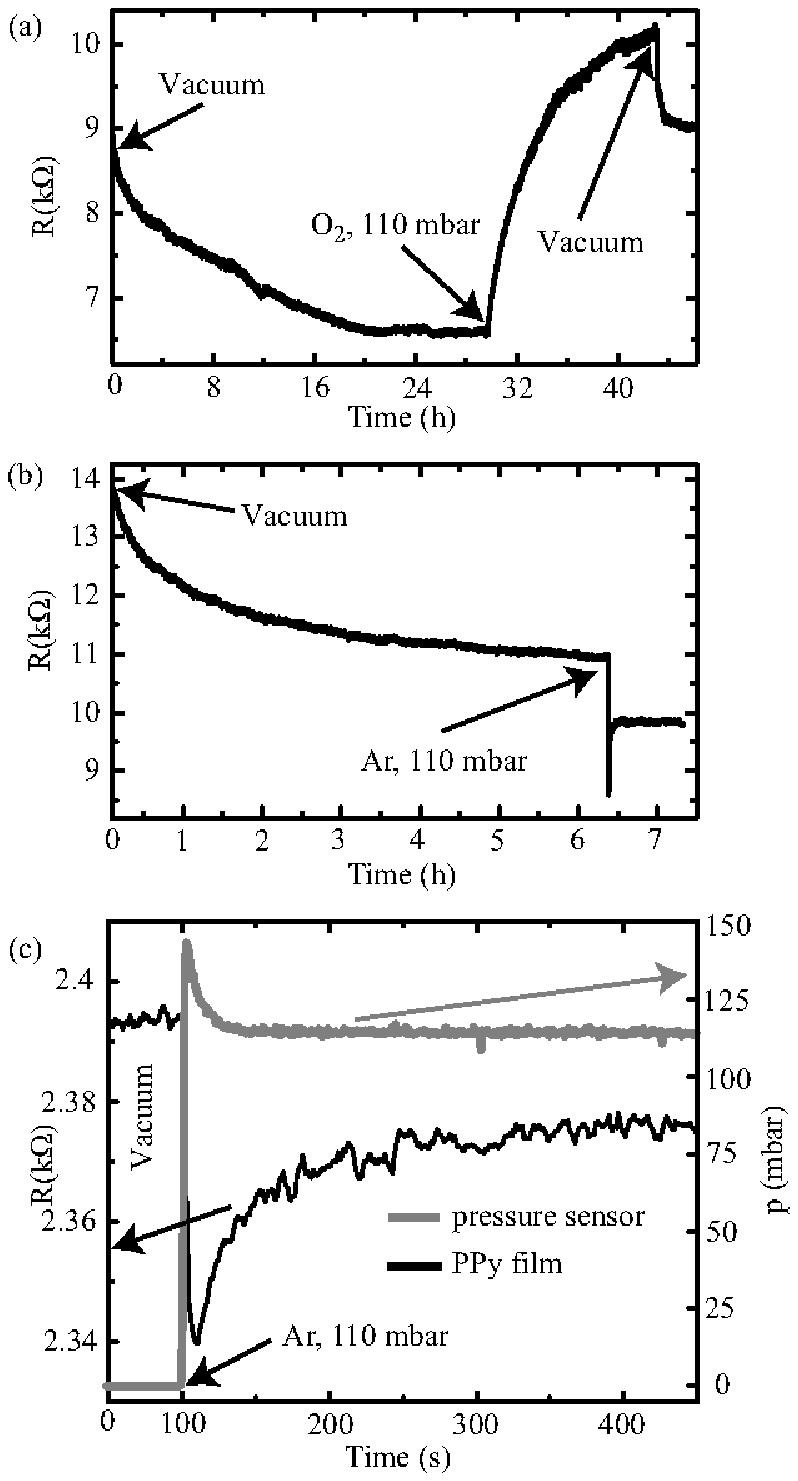}
\caption{}\label{PPy_Piezoresistivity_Fig2}
\end{figure}

In Fig. \ref{PPy_Piezoresistivity_Fig2}, the response of a PPy film of $50\,\mathrm{nm}$
thickness to the exposure of oxgen (a) and argon (b)is shown as a function of time.
Prior to gas exposure, the films are kept in vacuum, and a drop of the resistivity over time is
observed which saturates after about 1 day. We attribute this to vacuum
drying of  the film, which removes residues from the polymerization solution. As the film
is exposed to $150\,\mathrm{mbars}$ of oxygen, see Fig. \ref{PPy_Piezoresistivity_Fig2}(a), its resistance increases with the increase slowing down after about $12\,\mathrm{h}$, but does not show saturation even after 2 days. This response is only partly reversible when the vacuum is reestablished. Such a behavior is well
known from PPy films of larger thickness made by other techniques and is attributed
to overoxidation \cite{AlMashat2008a,Hamilton2005,Blanc1990,Lakard2007}.
Fig. \ref{PPy_Piezoresistivity_Fig2}(b) shows how this film responds to argon exposure. No chemical reaction is expected due to the inertness of the noble gas. However, application of $150\,\mathrm{mbars}$ of Ar causes the film resistance to drop and saturate within a few minutes. Since this response should have a purely
physical origin, we attribute it to a negative piezoresistance, $dR/dp <0$.
We note that the resistance dip observed immediately after exposure to $\mathrm{Ar}$ originates from the
pressure equilibration over time in the vacuum chamber. This is clearly demonstrated
in (c), where the film response is compared to that one of a commercial pressure sensor.
Apparently, the opening of the gas valve in our setup generates a pressure burst which
relaxes to the preset value within $\approx 30\,\mathrm{s}$. This comparison furthermore shows that
the response time of the PPy film is in the range of a few seconds.

\begin{figure}[htb]
\centering
\includegraphics{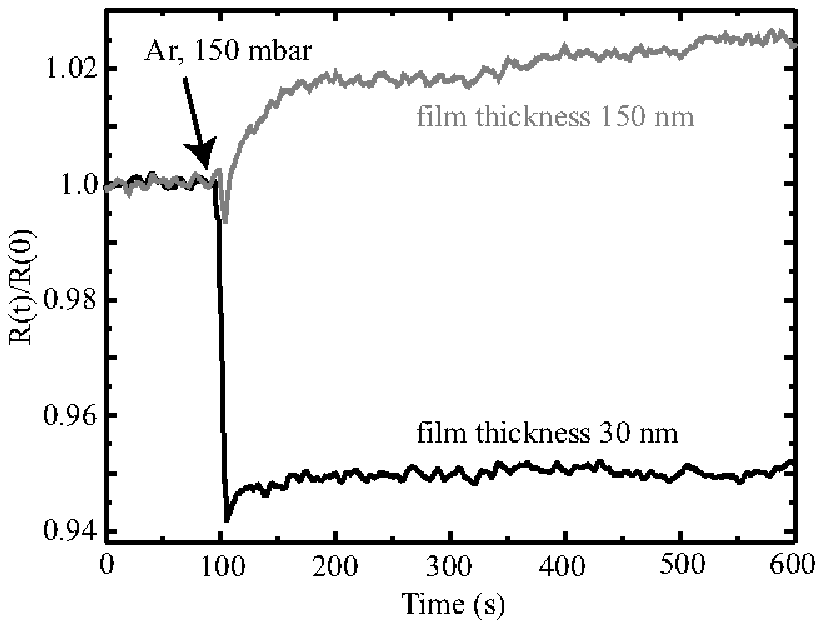}\caption{}
\label{PPy_Piezoresistivity_Fig3}
\end{figure}

In Fig. \ref{PPy_Piezoresistivity_Fig3}, the response to
$150\,\mathrm{mbars}$ of Ar pressure of  a PPy film with thickness $30\,\mathrm{nm}$
is compared to that one of  a film with thickness $150\,\mathrm{nm}$. While the PR is negative for the thin film, it is positive for the thicker film. After the pressure is applied, the resistance in the thin film drops by $\approx 5\%$ and saturates quickly. The resistance of the thicker film
increases by $\approx 1.5\%$ within one minute after pressurization, i.e., a significantly weaker response than the negative PR. It  keeps increasing slowly over a long period of time afterwards, with no clear signature of saturation even after 20 minutes. Note also that in the thicker film, a sharp spike towards lower resistance, i.e., a short term negative piezoresistive response, is observed immediately after the pressure is applied. This behavior, however, is quickly overcompensated by the positive PR just described.

\begin{figure}[htb]
\centering
\includegraphics{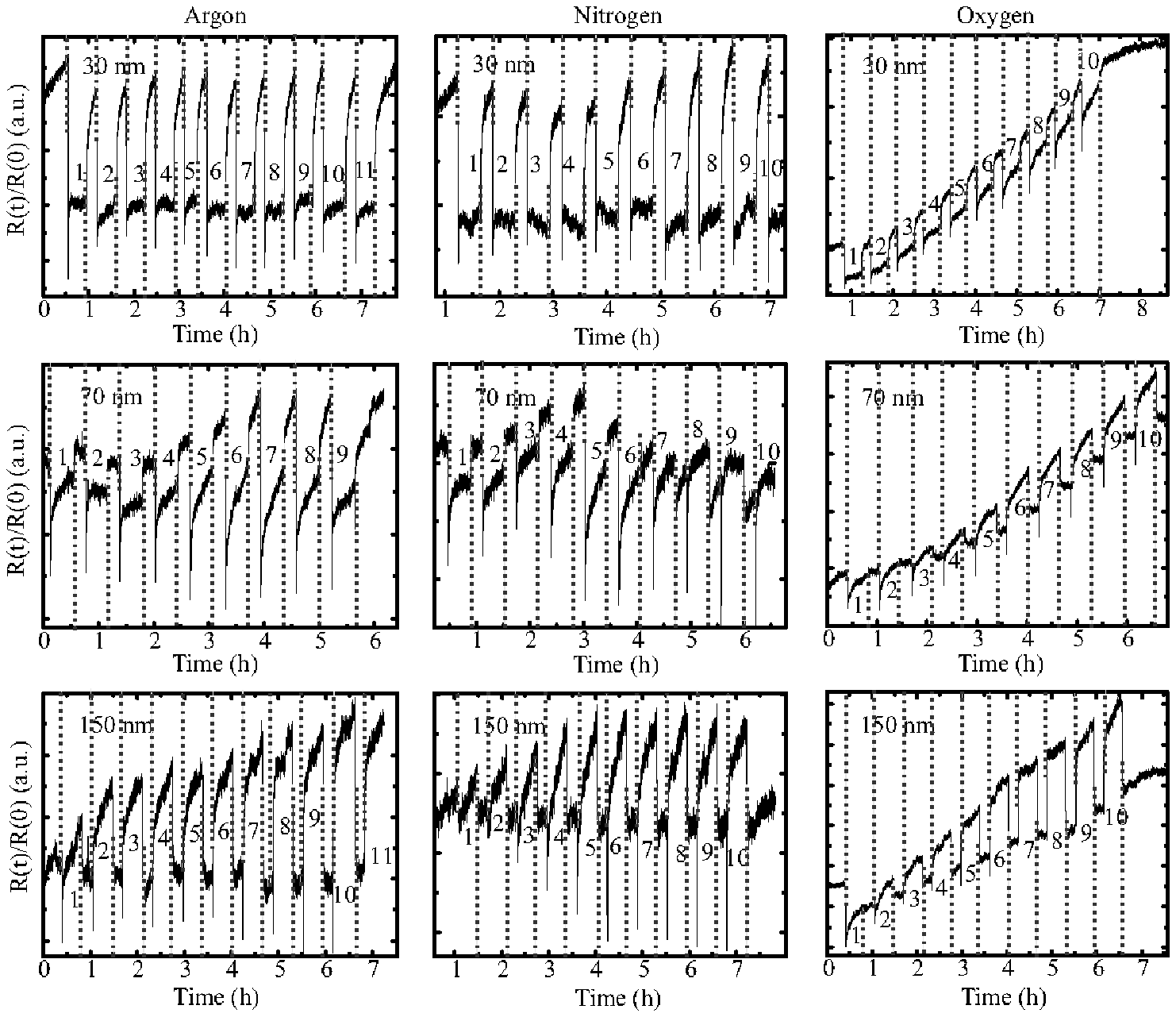}
\caption{}\label{PPy_Piezoresistivity_Fig4}
\end{figure}

The piezoresistivity studies were performed on films of thicknesses $30\,\mathrm{nm}$,
$50\,\mathrm{nm}$, $70\,\mathrm{nm}$, $100\,\mathrm{nm}$,
$120\,\mathrm{nm}$ and $150\,\mathrm{nm}$. An overview of the evolution of the dependence of the PR on film thickness, pressure and type of gas is given in Fig. \ref{PPy_Piezoresistivity_Fig4}. The negative (positive) PR is typical for all films of thicknesses well below (above) $\approx 70\,\mathrm{nm}$ at all pressures. At this critical thickness, a transition between the two responses takes place.

The films were exposed to a pressure which was stepwise increased between $50\,\mathrm{mbars}$ and $900\,\mathrm{mbars}$; in between the steps, vacuum was reestablished. The films were exposed to $\mathrm{Ar}$ and $\mathrm{N_2}$ as well as to $\mathrm{O_2}$ for comparison. The thin film (thickness $30\,\mathrm{nm}$) responded approximately identically to $\mathrm{Ar}$ and $\mathrm{N_2}$ pressure, and a pronounced negative PR is observed, with an approximately constant amplitude which does not show a
monotonous behavior as a function of the pressure. This effect is also observed in oxygen atmosphere. Here, however, we see also a slow but strong increase of the resistance which is persistent and most likely reflects the chemical response of the PPy film. This nicely demonstrates that the total response of the PPy films is a superposition of a physical and a chemical component. The thick film (thickness $150\,\mathrm{nm}$) shows a positive PR for all pressures, with the characteristic time dependence as discussed in Fig. \ref{PPy_Piezoresistivity_Fig3}. At an intermediate film thickness of $\approx 70\,\mathrm{nm}$, the positive and the negative piezoresistive components are of approximately equal strengths, which moreover depend slightly on the pressure. In the run with $\mathrm{Ar}$ - exposure, we observe that the time dependence of the resistance after the pressure exposure is different in comparison to the thin film. After the resistance drop, we no longer observe an approximately constant resistance; rather, it increases with a time dependence as observed in the thicker films with a positive PR, but does not exceed the resistance value under vacuum within our hold time of 25 minutes. This behavior gets more pronounced as the pressure is increased, which is in tune with the fact that the positive PR component has a stronger pressure dependence. In a second run, the same film was exposed to $\mathrm{N_2}$. Here, we observe a pressure- induced transition from negative to positive PR as the pressure is increased above $500\,\mathrm{mbars}$. Since the positive PR increases more strongly than the negative one as the pressure is increased, an overall switch of the PR polarity can be observed here. Note that a similar transition can be also seen in the experiment with oxygen. We remark that in our opinion, the differences observed between $\mathrm{Ar}$ and $\mathrm{N_2}$ exposure are not due to gas selectivity, but rather reflect the slowly changing properties of the film over time which are briefly discussed below.

Apparently, our PPy films do show a piezoresistive response to  inert gas atmospheres, which comprises two parts, namely a negative PR component that dominates in films of thicknesses below $70\,\mathrm{nm}$, saturates quickly and depends only weakly on the pressure level, and a positive PR component which dominates in films of thicknesses above $70\,\mathrm{nm}$, shows a stronger dependence on the pressure and saturates only on very long time scales of the order of hours. Furthermore, the negative PR component is also present in thicker films where the positive PR component dominates, as manifested in a sharp dip in the resistance as a function of time directly after pressurization. In contrast to this, the positive PR component is absent in the films which show a negative PR.

For a qualitative interpretation, we recall  (Fig. \ref{PPy_Piezoresistivity_Fig1}) that the film conductivity increases very strongly up to thicknesses of about $80\,\mathrm{nm}$ and depends only weakly on the film thickness for larger thicknesses. It appears plausible that for thicknesses below $\approx 80\,\mathrm{nm}$, the current-carrying states do not have bulk character but rather are localized surface or interface states with a two-dimensional character and a localization length that depends on the film thickness. Since the hopping conductivity depends exponentially on the localization length, \cite{BofBufon2007} a strong thickness dependence of the conductivity would result from this scenario. As the film thickness is increased beyond $80\,\mathrm{nm}$, the current is more and more carried by bulk states and the conductivity depends only weakly on the film thickness. Thus, it seems likely that the negative PR is related to a surface effect, while the positive PR originates from a bulk property of the film. Let us assume that the film is composed of a surface layer of $\approx 70\,\mathrm{nm}$, minus an unknown interface layer thickness due to the electronic structure of the film at the $\mathrm{SiO_x/PPy}$ interface, and a bulk layer in between the surface and the interface layer. Then, the surface-induced, negative PR would dominate for sufficiently thin films and also be present in thicker films, but would be of reduced relevance as the film thickness increases, due to the increasing weight of the positive bulk PR component. Suppose that the pressure just squeezes the surface layer somewhat. This would result in an increased overlap between the localized states responsible for the hopping transport, and hence in a reduced resistance. It can be expected that the squeezing occurs quickly and is stable once the pressure is applied. This is consistent with the observed fast and stable piezoresistive response. The mechanism responsible for the positive PR in the bulk must be strikingly different. We speculate that either pressure-induced energy shifts between the doping ions and the current carrying states result in a reduction of the hole density which in turn increase the resistance, or  that the gas molecules diffuse into the bulk and cause a film swelling which reduces the overlap between the current- carrying states, increases the hopping distance and thus increases the resistance. The fact that the positive PR develops over a long period of time suggests that diffusion of the gas molecules in the film plays an important role, which lets us favor the second scenario. Complementary studies are required to elucidate further these underlying mechanisms.

We conclude this Section with a remark regarding the long-term behavior of our samples and the significance of a quantification of the film response in terms of characteristic sensor quantities, like the sensitivity or the gauge factor. As suggested by the data shown in Fig. \ref{PPy_Piezoresistivity_Fig2}, the resistance of the films vary over time by as much as 50\% with typical time constants of hours. The origin of these changes are not completely clear. One mechanism is an initial drop of the resistance due to vacuum drying. Also, long-term memory effects are significant, like earlier exposure to oxygen or to inert gases of pressures above $\approx 500\,\mathrm{mbars}$. Another factor may be current- or voltage- induced conformational changes, well known from cyclic voltammetry performed on comparable films. \cite{BofBufon2005} The net effect of such mechanisms is a slowly varying time dependence of the resistance that is influenced by  the history of the sample and is not well understood. The piezoresistive sensitivity is defined as $s(p)\equiv \frac{R(p)-R(0)}{R(0)}$
where $R(p)$ denotes the resistance at a pressure $p$, and thus depends on the history of the sample. Consequently, this is also the case for the gauge factor, defined as $G\equiv ds(p)/dp$. Therefore, the values obtained for these quantities have to be considered with caution, and comparability is not given, not even for experiments performed on the same film at different times. Moreover, the resistance changes are not stable over time, in particular for the positive PR where saturation in reasonable periods of time is not observed, which means that the sensitivity depends strongly on the hold time after the application of the pressure. As a consequence, the absolute values are of little meaning and we therefore refrain from a quantitative discussion in terms of  $s$ or $G$. Rather, these considerations ask for a better understanding of the long-term behavior of the resistance as a prerequisite for a more quantitative analysis. Ultimately, applications of such films for accurate quantitative pressure or gas sensing will require an elimination of such drift effects.

\section{\label{sec:4}Summary and conclusions}

We have studied the response of the electronic transport in chemically synthesized
PPy films to exposure of different gases at various pressures. A pronounced piezoresistance is found which is composed of a negative and a positive component. In thin films with thicknesses below $70\,\mathrm{nm}$ , only the negative PR component is present. It  is characterized by a quick saturation in combination with a weak pressure dependence. In films of thicknesses $>70\,\mathrm{nm}$, a positive piezoresistance is dominant, while the negative contribution is also visible. This response does not show a clear saturation but rather evolves over many hours. However, its pressure dependence is stronger than the negative PR effect. Due to the different pressure dependencies of the two contributions, a pressure-induced transition from negative to positive PR is possible in films of $\approx 70\,\mathrm{nm}$ thickness. Attribution of the negative PR to a surface effect and the positive PR to a bulk effect is consistent with the observations. Moreover, measurements in oxygen atmospheres reveal that the chemical response used in sensing applications coexists with both kinds of piezoresistivities. At present, we have no indication that chemical sensitivity and the PR are interdependent.

Our results show that the response of polypyrrole films to gas exposure is more complicated than presently anticipated in the literature, and that piezoresistivity must be expected to be always present in addition to a chemical sensing mechanism. Thus, piezoresistivity may explain the ambiguous and nonmonotonous responses to gas exposures observed in earlier experiments, \cite{XingfaMa2008,Miasik1986,Hanawa1988} even though the PPy films were prepared by a different technology. Films with thicknesses below $70\,\mathrm{nm}$ respond unambiguously to the pressurization and the response saturates within seconds. Therefore, such thin PPy films appear better candidates for pressure sensors than thicker films where the bulk PR component becomes detectable as well.

Gas sensing applications typically require detection levels in the ppm regime and below. State of the art PPy films have shown resolution limits around 40 ppm \cite{Hernandez2007,Waghuley2008} for gas sensing and of about $500 \,\mathrm{N/m^2}$ in pressure sensing \cite{Brady2005}. It remains to be seen in future studies how this interplay between chemical and piezoresistive response scales into this regime. However, if we assume that the unusual behavior observed by Hanawa et al.\cite{Hanawa1988} on a 40 ppm level in $\mathrm{NO_2}$ - detection can be explained by piezoresistivity, the properties reported here will in fact be relevant in the interesting regime of detection levels. Moreover, we note that  oxygen sensing is important for pressure levels up to 1 atmosphere, where the results presented here are directly relevant. For example, a resistance measurement on a PPy thin film would not be able to distinguish between a pressure fluctuation and a change of the oxygen contents on a time scale of seconds. Finally, we note that the observed behavior also offers a potential solution to this ambiguity, for example by constructing resistance bridges with PPy arms of different thickness which nullify the piezoresistive signal.\\

Financial support by the Heinrich-Heine-Universit\"at D\"usseldorf is gratefully acknowledged.


\newpage

\noindent Author biographies:\\

\noindent  \emph{Said Barnoss}  made his M.Sc., in  Physics, at the Heinrich-Heine University of D\"{u}sseldorf in 2008. His research has been focused on the
sensing properties of polypyrrole during his master thesis.\\

\noindent  \emph{Dr. Hussein Shanak} got his PhD in Physics from Saarland University, Saarbr\"ucken, in 2005. He has worked in the field of thin film nanotechnology, especially transition metal oxides for electrochromic and gasochromic applications, and polyamide films for industrial applications. Since 2007 his research at D\"usseldorf University is focusing on transport in carbon -based materials including graphite systems and organic semiconductors.\\

\noindent  \emph{Dr. Carlos Cesar Bof Bufon} received his PhD in 2007 at the Heinrich-Heine University D\"usseldorf, where he investigated the transport properties of Polypyrrole films and devices.He is presently working as a postdoctoral researcher at the Institute for Integrative Nanosciences in Dresden (Germany) where he works on mechanoelectric silicon devices.\\

\noindent  \emph{Prof. Dr. Thomas Heinzel} made his PhD in Physics at the Ludwig.Maximilians-University in Munich (Germany) in 1994. He presently heads the Solid State Physics Laboratory at the  Heinrich-Heine University D\"usseldorf. His  research group is investigating mesoscopic transport in semiconductors.

\newpage
\noindent Figure captions:\\

\noindent  Figure 1: \\
\noindent Top: roughness and conductivity of the PPy films as a function of the film thickness. Bottom: scanning probe microscope image of the morphology of the film with $78\,\mathrm{nm}$ thickness used for this study (left). The roughness, i.e., the variance of the height fluctuations, of this film is $8.5\,\mathrm{nm}$. For comparison, the morphology of a film with a thickness of  $7.4\,\mathrm{nm}$ is shown to the right.\\

\noindent Figure 2:\\
\noindent Resistance of a 50 nm thick PPy film as a function of time when exposed
to oxygen (a) and argon (b). In (c), the time-resolved response of the film is shown
in comparison to the commercial pressure sensor used in the experiments. \\

\noindent Figure 3:\\
\noindent The normalized response of the film resistance to
exposure of $150\,\mathrm{mbars}$ Ar for PPy films of thickness $30\,\mathrm{nm}$
 with a negative piezoresistance in (a) and of $150\,\mathrm{nm}$ thickness
with a positive piezoresistance in (b).\\

\noindent Figure 4:\\
\noindent Piezoresistive response of films with thicknesses $30\,\mathrm{nm}$ (top), $70\,\mathrm{nm}$ (middle) and $150\,\mathrm{nm}$ (bottom), to exposure of $\mathrm{Ar}$ (left), $\mathrm{N_2}$ (center), and $\mathrm{O_2}$ (right) gas at various pressures, which are labeled as 1  to 4 for pressures from $50\,\mathrm{mbars}$ to $200\,\mathrm{mbars}$ in steps of $50\,\mathrm{mbars}$, and from 5 to 11 for pressures from $300\,\mathrm{mbars}$ to $900\,\mathrm{mbars}$ in steps of $100\,\mathrm{mbars}$, respectively. The dashed vertical lines from  the measurement traces  to the top and to the bottom indicate indicate the points of opening the gas inlet valve and its closing of with simultaneous begin of pumping to establish vacuum, respectively.

\newpage

\end{document}